\documentclass[doublecol]{epl2}
\usepackage{graphicx}
\date{Last revised   10.12.08}

\title{ 
 Comment on "Ergodicity and central-limit theorem in systems with  long-range  interactions" by Figueiredo et al.  \cite{Amato}  }
\shorttitle{Title} 

\author{A. Pluchino\inst{1} \and A. Rapisarda\inst{1} \and C. Tsallis\inst{2,3}}

\institute{
  \inst{1} Dip. di Fisica e Astronomia, Universit\`a di
Catania, and INFN - Via S. Sofia 64, I-95123
Catania, Italy\\
  \inst{2} Centro Brasileiro de Pesquisas Fisicas, - Rua Xavier Sigaud 150, 22290-180 Rio de Janeiro-RJ, Brazil \\
  \inst{3} Santa Fe Institute,  1399 Hyde Park Road, Santa Fe, NM 87501, USA
}


\abstract{}

\begin{document}

\maketitle

The Hamiltonian Mean Field  (HMF) model is a paradigmatic toy model for  long-range Hamiltonian 
system, which has raised recently a great attention and has been investigated in detail, see for example 
\cite{long-range} and references therein.
We  have recently published two papers  on this model \cite{hmf-clt1,hmf-clt2} concerning the possible application  of a generalized 
version of the Central  Limit Theorem (CLT) in accordance with $q$-statistics \cite{tsallis0,umarov,tirnakli,q-stat}.   
In these papers, following ref.\cite{tirnakli}, we constructed probability density functions (Pdf) of quantities $y$'s expressed as a finite sum of $n$ stochastic variables, selected along the deterministics time evolutions of the N rotors at fixed time intervals $\delta$.
We found that, along the metastable quasistationary states (QSS) of the HMF model , $q$-Gaussian-like  Pdfs appear to emerge as attractors \cite{hmf-clt1,hmf-clt2}.
In ref.  \cite{Amato} the authors advance several misleading claims  suggesting that 
the results    found  in   papers \cite{hmf-clt1,hmf-clt2} are erroneus and that no q-Gaussian attractors exist for correlated  variables  extracted  from HMF dynamics. 
Two  crucial points    raised  by Figuereido et al.  are the following: 
(i) by doing  the  calculations with more accuracy and better statistics,  it would be possible to show that  a q-Gaussian attractor does not exist,  since there is an evident cut-off in the tails of the final distributions;
(ii) by considering increasingly large size systems, the Pdf  would gradually converge to a Gaussian.  
The  point (i), which  is   illustrated  in fig.1 of   \cite{Amato},  can be easily rebutted  by 
  considering  that,  like the usual CLT, the $q$-CLT strictly applies only if  $n$ goes to infinity.
 In the example shown in  \cite{Amato} only  $n=50$ (with $N=100$ and $\delta=40$) was considered, 
 therefore  one expects   deviations in the tails, and elsewhere. 
Here  we present new calculations for the case $N=20000$, which indicate that this phenomenon is true whatever  is the shape of the final attractor. 
It is also important to remind that, in ref.\cite{hmf-clt2} (ref. [33] of \cite{Amato}), we observed the existence of three  classes of trajectories for the same water-bag  (M1) initial conditions, each one of them showing a different central limit behavior: class 1, with a $q$-Gaussian-like attractor,  class 3 with a Gaussian attractor and class 2 with an  intermediate behavior. 
In the upper row of fig.1 we show the Pdf of a typical event of class 1 for three increasing values of $n$ ($n=200$, $n=1000$ and $n=2000$), while in the middle row  the corresponding behavior  is illustrated for a typical event of class 3, for the same values of $n$. 
Being in all the simulations $\delta = 100$, the sums 
have been performed, respectively, for $20000$, $100000$ and $200000$ time steps. 
The  time evolution for the temperature in both cases is plotted  in the bottom panel.  
Notice that the cut-off in the tails of the distributions, 
clearly  visible for $n=200$,  disappears slowly by increasing the number of summands $n$, in both cases.  The convergence starts, for small $n$,  from the central part of the Pdf and then extends  to the tails, producing in one case (class 3) a Gaussian-like  and,  in the other case (class 1),  a  $q$-Gaussian-like curve as a final attractor.  
\begin{figure}
\begin{center}
\includegraphics [scale=0.33]
{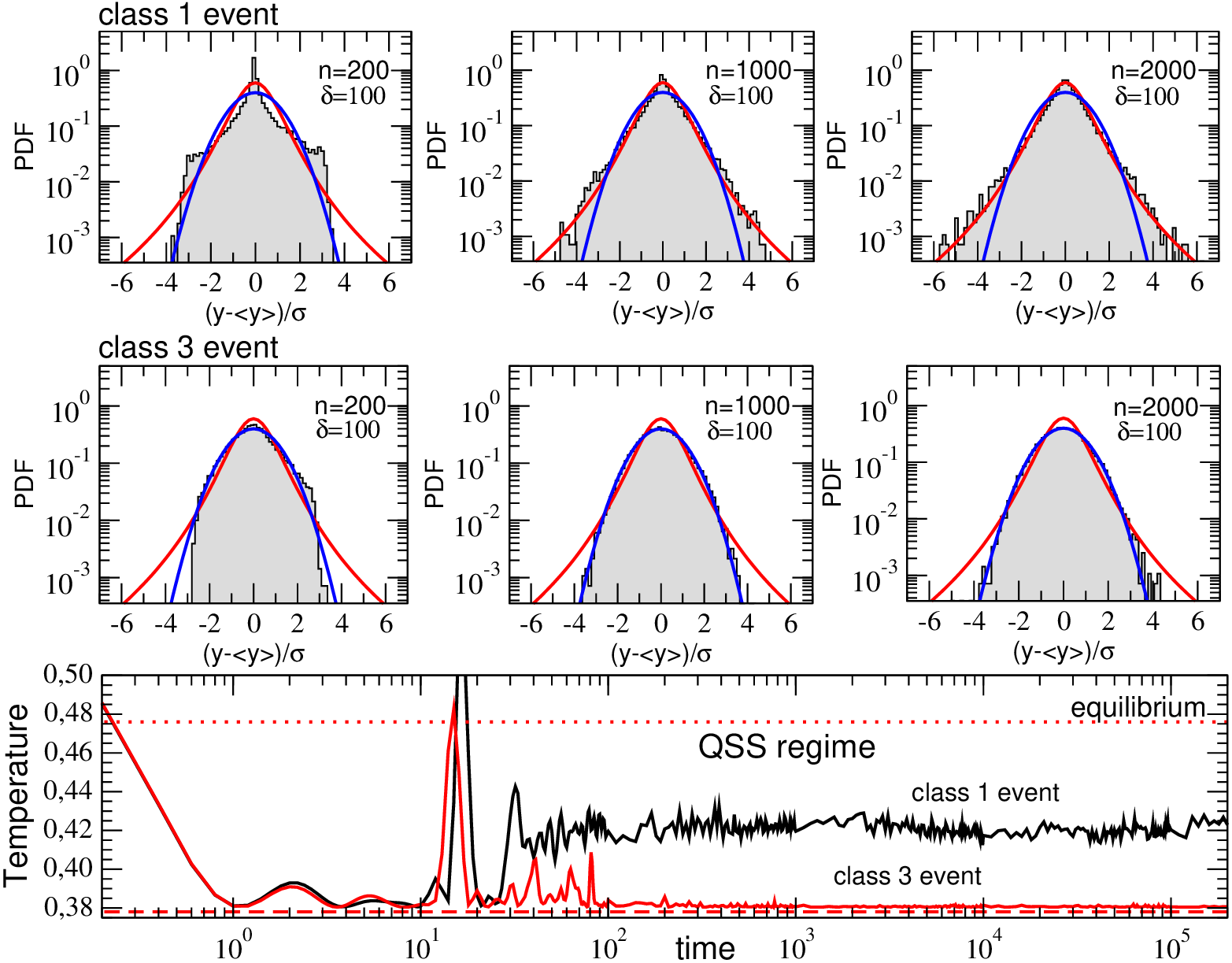}
\caption{Time evolution of two events with the same M1 initial conditions, more precisely
with $N=20000$ at $U=0.69$, but belonging to different classes (1 and 3). 
The  evolution  towards  the final attractor appears to be a Gaussian-like for the event of class 3 and a $q$-Gaussian-like for the event of class 1.  The $q$-Gaussian is a  generalization of Gaussian which emerges in the context of nonextensive statistical mechanics \cite{tsallis0} and is defined  as $ G_q(x)=A (1-(1-q) \beta x^2 )^{1/{1-q}} $. Here we have $q=1.42\pm0.1$,  
$\beta=1.3\pm0.1$ and $A=0.55$, and the resulting curve is plotted in all the panels toghether with a standard Gaussian. 
See text for further details.
} 
\label{fig1}
\end{center}
\end{figure}
%
\begin{figure}
\begin{center}
\includegraphics [scale=0.115] 
{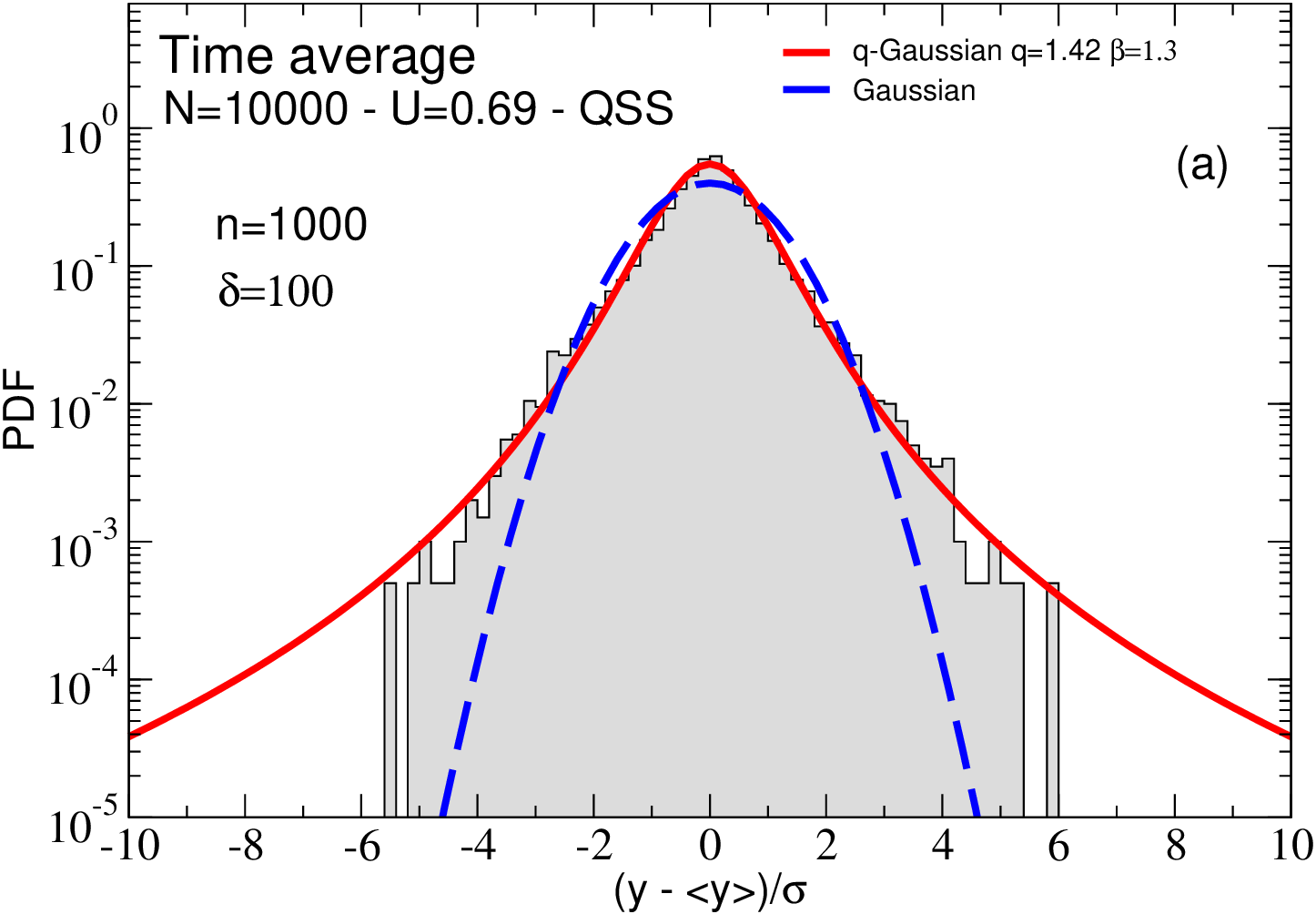}
\includegraphics [scale=0.115] 
{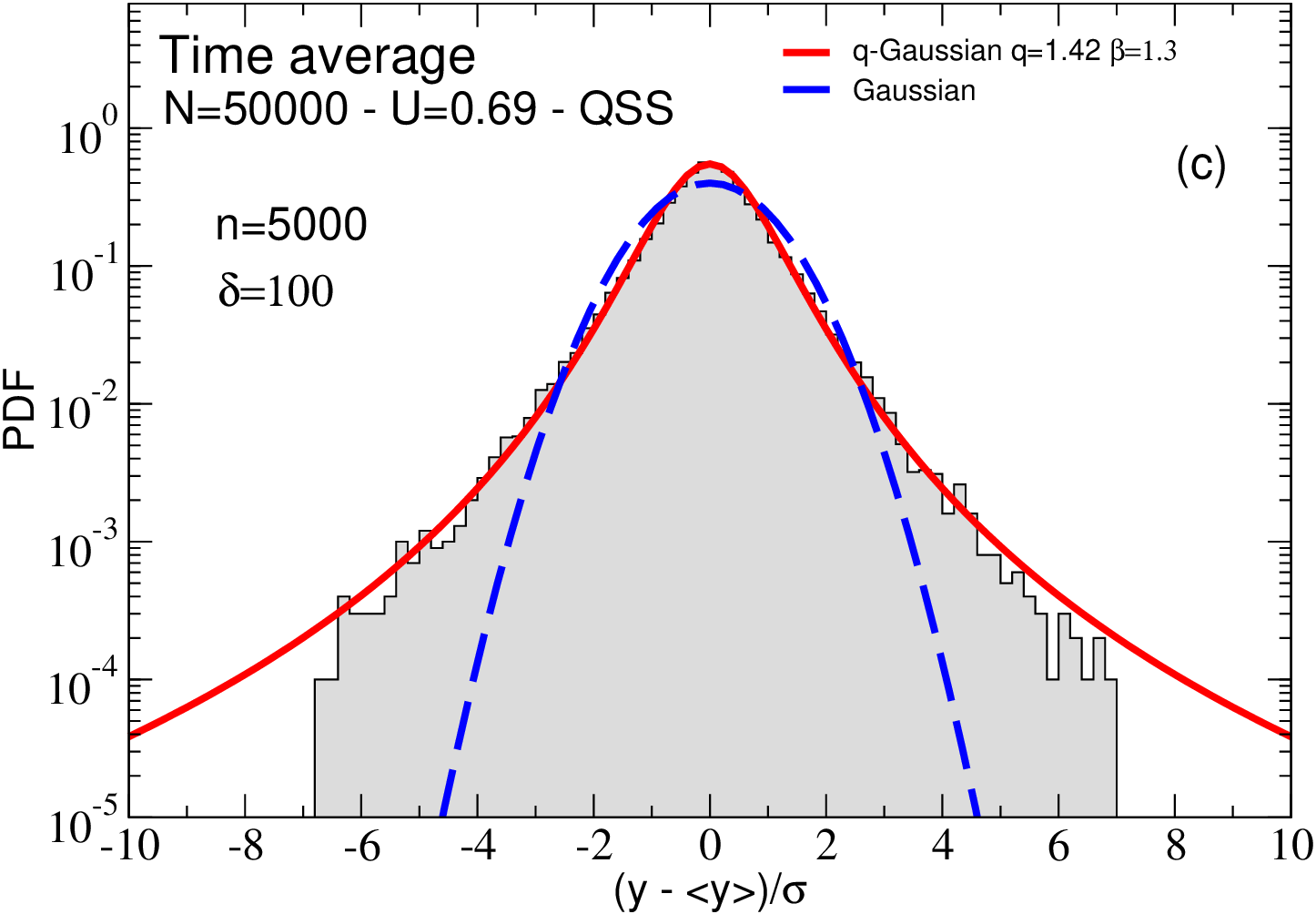}
\includegraphics [scale=0.115] 
{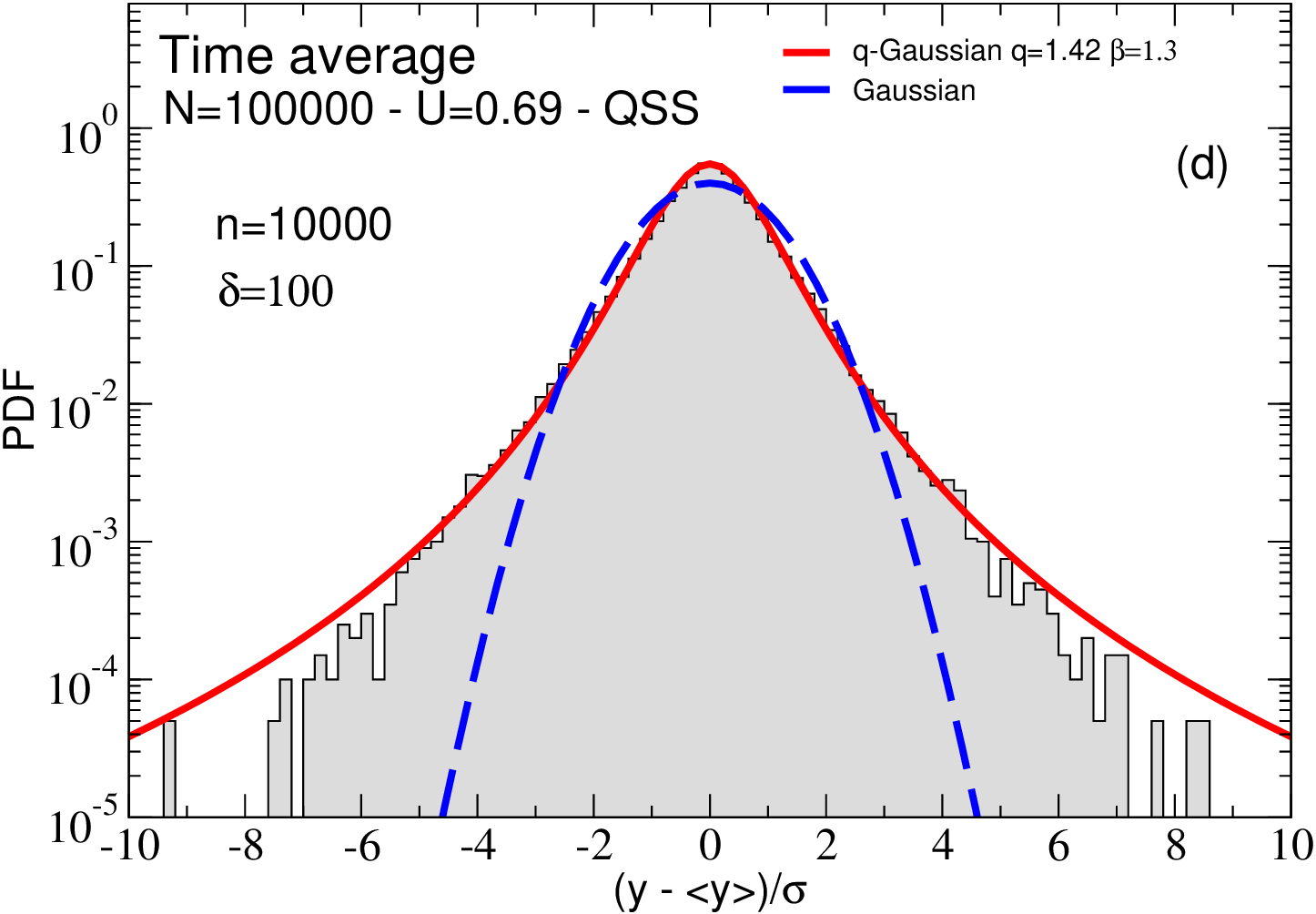}
\caption{ For $U=0.69$ we show the Pdfs obtained considering single events of class 1 for different sizes $N$, 
with $\delta=100$ and $n=N/10$. Again the indications for a $q$-Gaussian-like attractor  becomes stronger and stronger when  sending both $N $ and $n$ to infinity. 
 See text for further details.
} 
\label{fig2}
\end{center}
\end{figure}
%
The other main point addressed by fig.1 is that at least two kinds of attractors coexist in the QSS regime and their  occurrence,  as  discussed in ref. \cite{hmf-clt2}, depends  very sensitively on the different realizations of the initial conditions. 
The authors of ref. \cite{Amato}, although quoting our paper, fail in taking into account this fundamental point: 
in their fig. 3 they consider a system of $N=100000$ rotators and plot a Pdf for 50 events without any reference to their corresponding class, and it is evident that the trajectories used belonged to a random mixture of classes  1, 2 and 3 (see  \cite{hmf-clt2}). 
Actually, mixing events or not distinguishing between events of different classes leads to {unreliable} 
conclusions, since the dynamics and the correlations of the various trajectories can be very different. 
We notice that
although the events of class 1 seem to diminish with $N$, those of class 2 increase more than those of class 3 \cite{hmf-clt2}.
Finally, in order to address point (ii) of the  criticism raised in ref.\cite{Amato}, we  illustrate in fig. 2  how the attractor shown in the upper row of fig.1 evolves its shape by increasing both the size of the system $N$
 and, consistently, the number of summands $n$, but leaving constant the ratio $N/n$. 
 {This latter choice is  however not  a necessary condition, here it is done only because the QSS lifetime increases with $N$. }
 The calculations refer here to three typical events
 of class 1 with sizes $N=10000,50000$ and $100000$ (note that the range of the ordinate was enlarged with respect to fig. 1). The gradual filling of the tails by increasing $N$ strongly suggests that, for $N$ and $n$  going to infinity, a $q$-Gaussian-like asymptotic shape might well be eventually reached, thus reinforcing its role of real attractor for the QSS regime.
In conclusion,  the evidence   presented in \cite{Amato}  by no means invalidates the occurrence of  the $q$-generalized  CLT for the HMF system and in general for long-range Hamiltonian systems. 
{On the other hand,  if these systems  are non-ergodic (as the authors themselves admit and as we believe we have clearly established \cite{hmf-clt1,hmf-clt2}),  it also remains unclear why should one not expect, for the time-averaged distribution, a non-Gaussian attractor in the QSS  regime.  }
In summary, we presented further numerical evidence that  standard  statistical mechanics has  severe  problems in explaining the anomalous  long-standing QSS regime of the HMF model, showing that 
in this respect $q$-statistics remains  a very good candidate.


\begin{thebibliography}{0}
\bibitem{Amato}
  \Name{Figueiredo A., Rocha Filho T.M., Amato M.A.}
  \REVIEW{Europh. Lett.} {83}{2008} {30011}

\bibitem{long-range}
 \Name{Dauxois T., Latora V. , Rapisarda A. , Ruffo S. \and Torcini A.}
\Book{ Lecture Notes in Physics}
\Editor{T. Dauxois, S. Ruffo, E. Arimondo, M. Wilkens}
\Vol{602} \Year{2002} \Page{458};
%
 \Name{Latora V., Rapisarda A.  \and Ruffo S.}
  \REVIEW{Phys. Rev. Lett.} {80} {1998} {692};
%
 \Name{Latora V., Rapisarda A.  \and Tsallis  C.}
  \REVIEW{Phys. Rev. E} {64} {2001} {056134};
%
 \Name{Pluchino A.,  Latora V.  \and Rapisarda A.}
  \REVIEW{Physica D} {193} {2004} {315};
%
 \Name{Rapisarda A. \and Pluchino A.}
  \REVIEW{Europhysics News} {36} {2005} {202};
%
%
  \Name{Giansanti A., Moroni D.  \and Campa  A.}
  \REVIEW{Physica A}{305}{2002}{137} ;
    \REVIEW{J. Phys. A }{36}{2003}{6897} ;
%
%
  \Name{ Chavanis P-H.}
  \REVIEW{Eur. J. Phys. B}{53}{2006}{487};
%
  \Name{Antoniazzi A., Califano F., Fanelli D.,  \and Ruffo S.}
  \REVIEW{Phys. Rev. Lett}{98}{2007}{150202};
%
  \Name{Morita H.  \and Kaneko K.}
  \REVIEW{Phys. Rev. Lett.}{96}{2006}{050602};
%
%
  \bibitem{hmf-clt1}
  \Name{Pluchino  A. ,  Rapisarda A. \and Tsallis  C.}
  \REVIEW{Europhysics Letters}{80}{2007}{26002}

 \bibitem{hmf-clt2}
   \Name{Pluchino  A. ,  Rapisarda A. \and Tsallis  C.}
  \REVIEW{Physica A}{387}{2008}{3121}
  
\bibitem{tsallis0}
  \Name{Tsallis  C.}  
  \REVIEW{J. Stat. Phys.}{52}{1988}{479};
%
  \Name{Tsallis C.,  Gell-Mann M. \and Sato Y.}
  \REVIEW{Europhys. News}{36}{2006}{186}, and references therein;
%
  \Name{Tsallis C.}
   \REVIEW{Milan J. Mathematics} {73}{2005}{145}, and references therein

 \bibitem{umarov}
   \Name{S. Umarov, C. Tsallis and S. Steinberg}
   \REVIEW{cond-mat/0603593, Milan J. Math.}{76}{2008}{DOI 10.1007/s00032-008-0087-y}

\bibitem{tirnakli}
  \Name{Tirnakli U., Beck C. \and Tsallis  C.}
  \REVIEW{Phys. Rev. E}{75}{2007}{040106 (R)};
%
  \Name{Tirnakli U., Tsallis C., Beck  C.}
 \REVIEW{0802.1138, cond-mat.stat-mech} {}{2008}{}

\bibitem{q-stat}
 \Name{F. Baldovin F. \and A. Stella}
 \REVIEW{Phys Rev. E} {75} {2007} {020101(R)};
%
 \Name{Moyano L.G., Tsallis C., Gell-Mann M.}
 \REVIEW{Europhys. Lett.} { 73} {2006}{813};
%
 \Name{Thistleton  W.J., Marsh J.A., Nelson K.  \and Tsallis C.}
 \REVIEW{IEEE Transactions on Information Theory } {53}{2007}{4805};
%
 \Name{Hilhorst H.J.  \and Schehr G.}
  \REVIEW{JSTAT} {06} {2007} {P06003};
%
 \Name{Vignat C. \and Plastino A.}
 \REVIEW{J. Phys. A } {40}{2007}{F969};
 \Name{Rodriguez A., Schwammle V., Tsallis C.}
  \REVIEW{0804.1488 [cond-mat.stat-mech] JSTAT} {2008}{ in press}




\end{thebibliography}
\end{document}